\documentclass[runningheads]{llncs}
\usepackage{graphicx}
\usepackage{amsmath}
\usepackage{paralist}

\usepackage{todonotes}

\newcommand{\bim}[2][]{\mathop{\mathrm{RSJ}^{#1}_{#2}}}
\newcommand{\rsj}[2][]{\mathop{\mathrm{RSJ}^{#1}_{#2}}}

\newcommand{\deltaRSJ}{\ensuremath{\Delta\mathrm{RSJ}}}

\begin{document}
\title{Match Your Words! A Study of Lexical Matching in Neural Information Retrieval
}
\titlerunning{Match your words!}
%
\author{Thibault Formal\inst{1,2} \and
Benjamin Piwowarski\inst{2,3} \and
Stéphane Clinchant\inst{1}}


%
\authorrunning{T. Formal et al.}

%
\institute{Naver Labs Europe, Meylan, France \\ \email{\{thibault.formal,stephane.clinchant\}@naverlabs.com} \and
Sorbonne Université, LIP6, Paris, France\\
\email{benjamin.piwowarski@lip6.fr}
\and CNRS}

\maketitle              
\begin{abstract}

Neural Information Retrieval models hold the promise to replace lexical matching models, e.g. BM25, in modern search engines. While their capabilities have fully shone on in-domain datasets like MS MARCO, they have recently been challenged on out-of-domain zero-shot settings (BEIR benchmark), questioning their actual generalization capabilities compared to bag-of-words approaches. Particularly, we wonder if these shortcomings could (partly) be the consequence of the inability of neural IR models to perform lexical matching off-the-shelf. In this work, we propose a measure of discrepancy between the lexical matching performed by any (neural) model and an ``ideal'' one. Based on this, we study the behavior of different state-of-the-art neural IR models, focusing on whether they are able to perform lexical matching \emph{when it’s actually useful}, i.e. for important terms. Overall, we show that neural IR models fail to properly generalize term importance on out-of-domain collections or terms almost unseen during training.

\end{abstract}

\section{Introduction}


Over the last two years, the effectiveness of neural IR systems has risen substantially. Neural retrievers based on pre-trained Language Models like BERT~\cite{DBLP:conf/naacl/DevlinCLT19} -- whether dense or sparse -- hold the promise to replace lexical matching models (e.g. BM25) for first-stage ranking in modern search engines. 
%
Despite this success, little is known regarding their actual inner working in the IR setting. Previous works scrutinizing BERT-based ranking models either relied on axiomatic approaches adapted to neural models~\cite{CamaraDiagnosingBERTRetrieval2020,sciavolino2021simple}, controlled experiments~\cite{MacAvaneyABNIRMLAnalyzingBehavior2020}, or direct investigation of the learned representations~\cite{JiangHowDoesBERT2021,FormalWhiteBoxAnalysis2021} or attention~\cite{10.1145/3437963.3441667}. 
This line of work has shown -- among other findings -- that these models, which rely on contextualized semantic matching, are actually still quite sensitive to lexical match and term statistics in documents/collections~\cite{JiangHowDoesBERT2021,FormalWhiteBoxAnalysis2021}. However, these observations are based on specifically tailored approaches that cannot directly be applied to any given model. To generalize these findings, we introduce instead an intuitive black box approach: we propose to ``count'' query terms appearing in top documents retrieved by various state-of-the-art neural systems, in order to compare their ability to perform \emph{lexical matching}.


%

Furthermore, previous studies have been conducted on the MS MARCO dataset, on which models have been trained. The BEIR benchmark~\cite{ThakurBEIRHeterogenousBenchmark2021a} has shown that the only systems improving the overall performance over BM25 in the zero-shot setting have (somehow) a lexical bias, e.g. models like doc2query-T5~\cite{Nogueiradoc2querydocTTTTTquery} or ColBERT~\cite{KhattabColBERTEfficientEffective2020}. Therefore, we also propose to study the extent to which neural IR models are able to \emph{generalize} lexical matching, for query terms that either have not been seen in the training set or with different collection statistics (e.g. common in the training set but rare on an out-of-domain evaluation set).
%
%

In this work, we first develop indicators that help measuring to what extent a lexical match is ``important'' for the user (user relevance) or for the model (system relevance). 
By comparing both values -- i.e. computing the difference between the user and the system, we can look at the following research questions:
\begin{inparadesc}
\item[(RQ1)] To what extent neural retrievers perform accurate lexical matching (Sect. \ref{sec:lexical-match})?
\item[(RQ2)] Do they generalize term matching to unseen query terms (Sect. \ref{sec:lexical-match})?
\item[(RQ3)] Do they generalize term matching to new collections (Sect. \ref{sec:generalization:zero-shot})?
\end{inparadesc}

\section{Methodology}\label{sec:methodo}
Our analysis rationale is the following: the more a term is important for a query (w.r.t. relevant documents), the more frequent the term should be retrieved by the system in top retrieved documents. Therefore, we first need to define what it means for a term to be \emph{important for lexical matching},  and how to accurately measure frequency in top documents. Roughly speaking, we are interested in the models ability to retrieve documents containing query terms, \emph{when they are deemed important}. Note that we are not interested in expansion mechanisms in our analysis since they are more related to semantic matching.

Intuitively, term importance w.r.t. relevance can be measured by the extent to which a term allows to distinguish relevant from non-relevant documents in a collection of documents.
It is thus natural to use the Robertson-Sparck Jones (RSJ) weight~\cite{YuPrecisionWeightingEffective1976,RobertsonRelevanceweightingsearch1976}. The RSJ weights have been shown, if estimated correctly, to order documents in the optimal order w.r.t. the Probability Ranking Principle~\cite{RobertsonProbabilityRankingPrinciple1977}. 
For a given user information need $U$, the user $\rsj{U}$ weight for term $t$ is defined as follows (the conditioning on query $q$ is implicit):
\begin{equation}
\bim{t,U} =
    \log \frac 
    {p(t|R) p(\neg t| \neg R)}
    {p(\neg t|R) p(t| \neg R)}
\label{eq:bim}    
\end{equation}
where $P(t|R)$ is the probability that term $t$ occurs in a relevant document. $\bim{t,U}$ is thus high when a term, for a document to be relevant, is both  \emph{necessary} ($p(.|R)$) and \emph{sufficient} ($p(.|\neg R)$). Intuitively, it is low for e.g. stopwords, as they have equal \emph{odds} to appear in relevant and irrelevant documents. 
The above weight can be estimated using the set of relevant documents and collection statistics.

We now want to compute the same weight, when relevance is defined by the \emph{system} (and not the \emph{user}). In other words, we would like to measure how much a model ``retrieves'' term $t$. One way to proceed is to suppose that top-$K$ documents are \emph{relevant from the point of view of the system}, for a suitable $K$. 
While a more accurate definition of system relevance could be used, we found out in our preliminary analysis that results were not very sensitive to the choice of $K$.
%
We hence define the system $\rsj{S}$ weight for term $t$ as:
\begin{equation}
\bim{t,S} = 
    \log \frac 
    {p(t|\text{top-}K) p(\neg t| \neg \text{top-}K)}
    {p(\neg t|\text{top-}K) p(t| \neg \text{top-}K)}
\label{eq:bim_S}    
\end{equation}

Intuitively, it gives us a mean to properly count occurrences of query terms in retrieved documents -- taking into account collection statistics. It is estimated similarly to Eq.~\ref{eq:bim}.
Once $\rsj{U}$ and $\rsj{S}$ have been computed, we can look at the difference between both, i.e. $\deltaRSJ_t = \rsj{t,S} - \rsj{t,U}$. If $\deltaRSJ_t > 0$ (resp. $\deltaRSJ_t < 0$), it means that the model overestimates (resp. underestimates) the importance of term $t$ when considering its document ordering. In other words, the model retrieves ``too much'' (resp. ``too few'') this term.
%
%
Please note that a high correlation between $\rsj{S}$ and $\rsj U$ \textbf{is not} indicative of the absolute performance of a model, as $\rsj U$ is neither a perfect model nor performance measure. However, we argue that it can still indicate partly the performance of the model w.r.t. lexical matching, especially for terms whose $\rsj U$ are high.


\section{Experiments}


We conducted experiments by analyzing models trained on MS MARCO~\cite{journals/corr/NguyenRSGTMD16}, using public model parameters when available (indicated by $\star$). We evaluated models on the \emph{in-domain} TREC Deep Learning 2019-2020 datasets~\cite{craswell2020overview,craswell2021overview} ($97$ queries in total), and two \emph{out-of-domain} datasets from the BEIR~\cite{ThakurBEIRHeterogenousBenchmark2021a} benchmark (\texttt{TREC-COVID} (bio-medical) and \texttt{FiQA-2018} (financial), with respectively $50$ and $648$ test queries).
For all our experiments, we measure the system relevance by using top-$K=100$.
For the term-level analysis, we keep stopwords, and use standard tokenization and Porter stemming. We solely focus on first-stage retrievers (and not re-rankers), for which lexical matching might be more critical. We thus compare various state-of-the-art models (based on the BEIR benchmark), considering different types of approaches (sparse and dense). We include two lexical models, the standard \texttt{BM25}~\cite{RobertsonProbabilisticRelevanceFramework2009} and \texttt{doc2query-T5} ($\star$)~\cite{Nogueiradoc2querydocTTTTTquery}; \texttt{SPLADE} ($\star$)~\cite{10.1145/3404835.3463098,FormalSPLADEv2Sparse2021}, an expansion-based sparse approach; \texttt{ColBERT}~\cite{KhattabColBERTEfficientEffective2020}, an interaction-based architecture; two dense retrievers, \texttt{TAS-B} ($\star$)~\cite{HofstatterEfficientlyTeachingEffective2021} and a standard \texttt{Bi-encoder} trained with contrastive loss and in-batch negatives.


\subsection{Lexical Match in Neural IR}
\label{sec:lexical-match}

\begin{figure}
    \centering
    \includegraphics[width=1\textwidth]{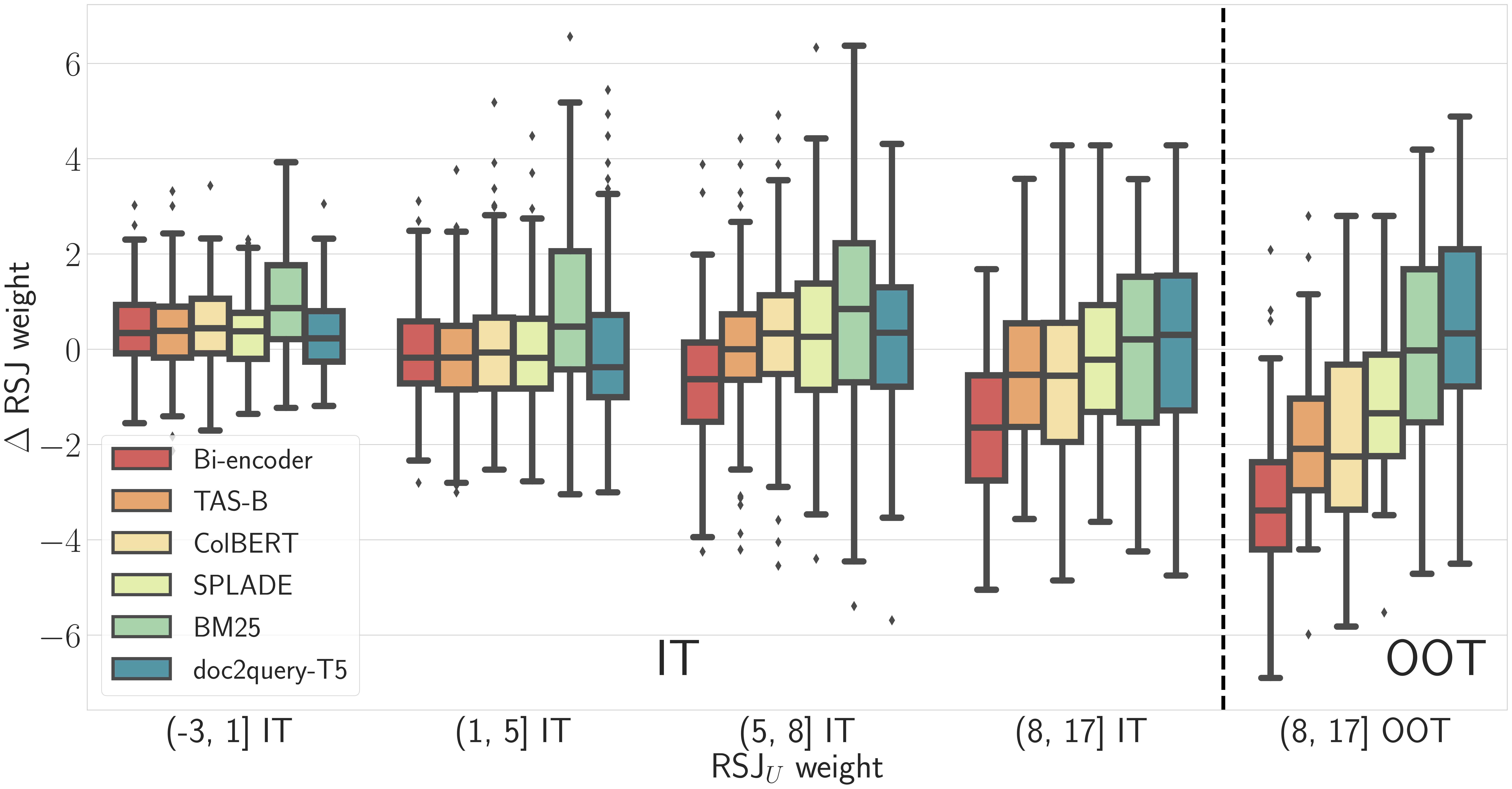}
    \caption{$\deltaRSJ$ with respect to user $\rsj{U}$ (x-axis, binned), splitting according to query terms seen during training (\texttt{IT}, left) or not (\texttt{OOT}, right). We consider that terms appearing in less than $10$ training queries are \texttt{OOT}, leading to $499$ and $42$ terms in TREC queries, for \texttt{IT} and \texttt{OOT} respectively. Note that due to the fact that \texttt{OOT} terms are also generally rare in the collection, their $\rsj{U}$ is always $>8$, hence the single bin.}
    \label{fig:qr1-bim-all}
\end{figure}



In Fig.~\ref{fig:qr1-bim-all}, we plot the relationship between the user weight and $\deltaRSJ$, for each term in the test queries appearing at least 10 times in the training queries (left, \texttt{IT} for In-Training). We first note that lexical-based models tend to overestimate the importance of query terms ($\deltaRSJ >0$). The second observation is that models are roughly similar in their estimations for low user $\rsj U$ weights (below 5).
Then, there is a clear distinction between the bi-encoder and other neural models (both dense and sparse): we can see that it retrieves less documents, on average,  containing precisely the important query terms.
Comparing dense and sparse/interaction models overall -- by considering the average $\deltaRSJ$ over terms -- we observe that, interestingly, dense models underestimate $\rsj U$ ($\overline \deltaRSJ=-0.07$ for TAS-B and $-0.26$ for the bi-encoder) while sparse/interaction slightly 
overestimate it ($\overline \deltaRSJ=0.03$ for ColBERT and SPLADE). Note again, as mentioned in Sect.~\ref{sec:methodo}, that the measure is not necessarily indicative of performance: for instance, TAS-B performs better than BM25 on TREC, suggesting that the model is better for semantic search. To illustrate the above, let us consider a query from the TREC DL set: "\texttt{does} (-1.12) \texttt{legionella} (14.85) \texttt{pneumophila} (13.12) \texttt{cause} (4.34) \texttt{pneumonia} (8.34)" (terms with associated $\rsj U$). BM25 is able to correctly estimate importance for \texttt{legionella} ($\rsj S=15.08$) contrary to neural approaches which tend to under-estimate it ($\rsj S=10.63,13.42,13.65$ for the bi-encoder, SPLADE and ColBERT respectively).



We now shift our attention to the behavior of models for query words that are \emph{(almost) not} in the training set.
In Fig.~\ref{fig:qr1-bim-all}, we show the distribution of $\deltaRSJ$ for terms appearing in less than $10$ training queries (out of $> 500k$) (right, \texttt{OOT} for Out-Of-Training). Comparing with $\deltaRSJ$ for terms in the training set, we can see that all neural models are affected somehow, showing that lexical match does not fully generalize to ``new'' terms. For the $(8,17]$ bin, and for every model (except BM25), the difference in mean between \texttt{IT}/\texttt{OOT} is significant, based on a $t$-test with $p=0.01$. 

\begin{figure}[t!]
    \centering
    \includegraphics[width=0.9\textwidth]{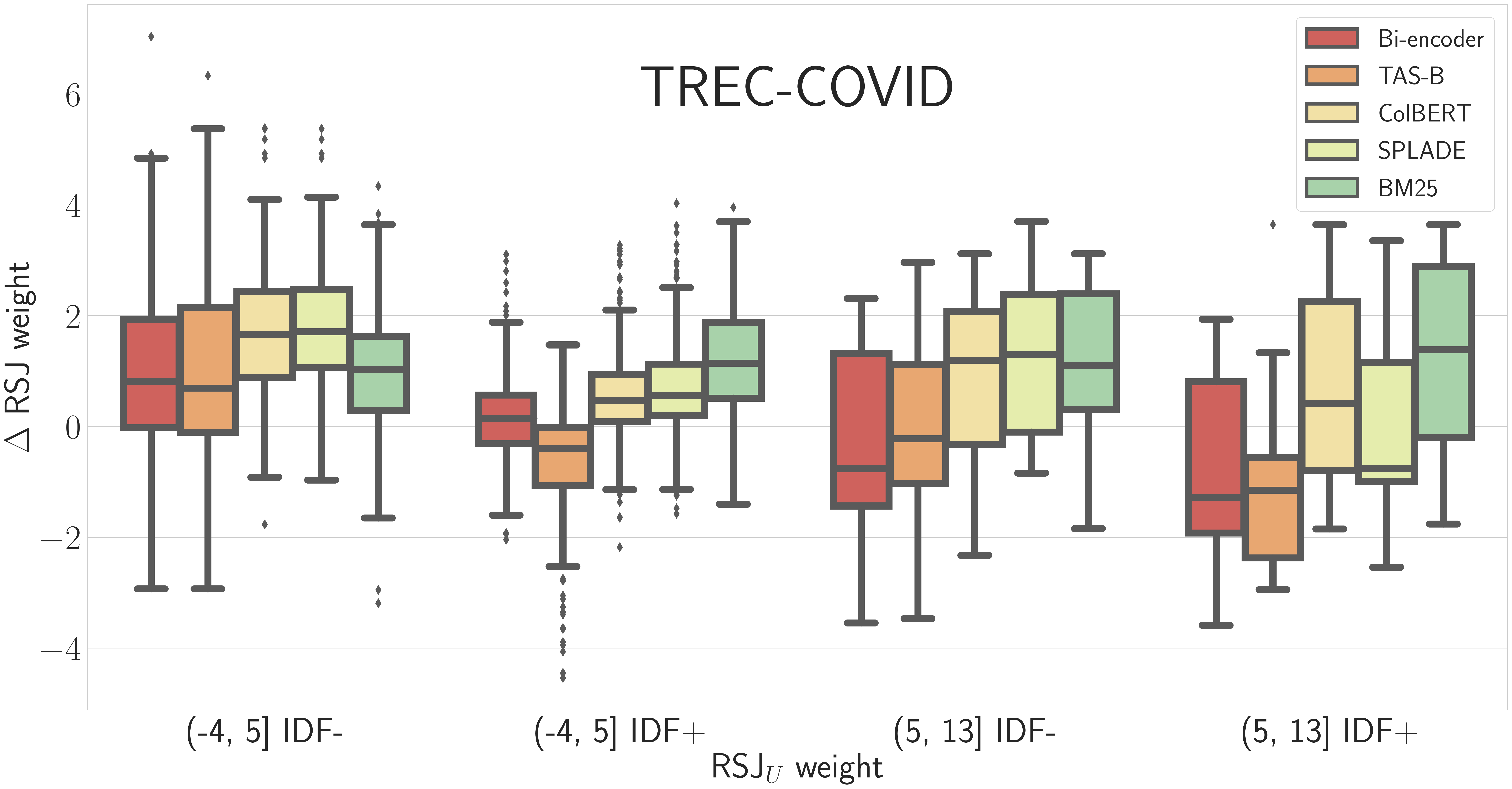}\\
    \includegraphics[width=0.9\textwidth]{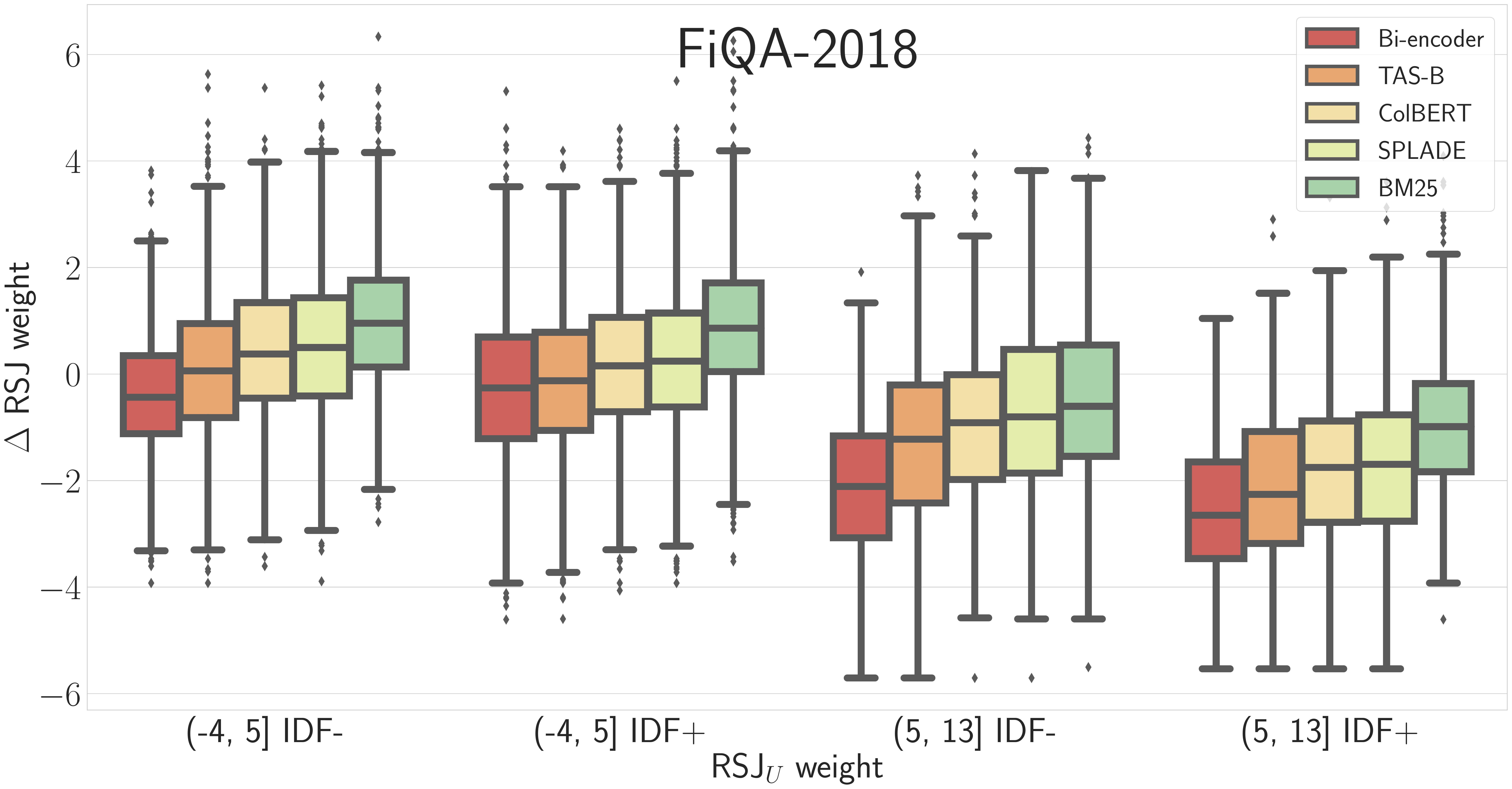}
    \caption{$\deltaRSJ$ with respect to $\rsj{U}$ (x-axis, binned) in the zero-shot setting. \texttt{IDF-} includes $108$ and $933$ terms, while \texttt{IDF+} includes $112$ and $428$ terms for respectively \texttt{TREC-COVID} and \texttt{FiQA-2018}. Note that bins are not similar compared to Fig.~\ref{fig:qr1-bim-all}, as RSJ  weights have different distributions on BEIR datasets.}
    \label{fig:qr3-zero-shot}
\end{figure}

Finally, we also looked at the relationship between \texttt{IT}/\texttt{OOT} and  model performance. More precisely, for terms in the $(8,17]$ bin, we computed the mean ndcg@10 for queries containing at least one term either in \texttt{IT} or \texttt{OOT} (respectively $55$ and $37$ queries out of the $97$, with $9$ queries in both sets). We found that BM25 and doc2query-T5 performance increased by $0.1$ and $0.02$ respectively, while for all neural models the performance decreased ($\approx 0$ for TAS-B, -0.11 for SPLADE, -0.27 for the bi-encoder and -0.38 for ColBERT). The fact that BM25 performance increased is likely due to the fact that the mean IDF increased (from $7.3$ to $10.9$), i.e. important terms are more discriminative in the \texttt{OOT} query set. 
With this in mind, the decrease of all neural models might suggest that a potential reason for the relative performance decrease (w.r.t. BM25) is due to a worse estimate of high $\rsj{U}$.

\subsection{Lexical Match and Zero-Shot Transfer Learning}
\label{sec:generalization:zero-shot}

We now analyze whether term importance can generalize to the zero-shot setting\footnote{We excluded doc2query-T5 from the analysis, due to the high computation cost for obtaining the expanded collections.}. 
We distinguish two categories of words, namely those that occurred $5$ times more in the target collection than in MS MARCO (\texttt{IDF+}), or those for which term statistics were more preserved (\texttt{IDF-}), allowing us to split query terms in sets of roughly equal size. Since term importance is related to the collection frequency (albeit loosely), we can compare $\deltaRSJ$ in those two settings. Fig.~\ref{fig:qr3-zero-shot} shows the $\deltaRSJ$ with respect to $\rsj{U}$ for the \texttt{TREC-COVID} and \texttt{FiQA-2018} collections from the BEIR benchmark~\cite{ThakurBEIRHeterogenousBenchmark2021a}.

We can first observe that neural models underestimate $\rsj{U}$ for terms that are more frequent in the target collection than in the training one (\texttt{IDF+}). It might indicate that models have learned a dataset-specific term importance -- confirming the results obtained in the previous section on out-of-training terms. 
When comparing dense and sparse/interaction models overall -- by considering the average $\deltaRSJ$ over terms -- we observe than dense models underestimate even more $\rsj{U}$ than on in-domain ($\overline{\deltaRSJ}=-0.17$ for TAS-B and $-0.38$ for the bi-encoder) while sparse/interaction seem to overestimate ($\overline{\deltaRSJ}=0.18$ for ColBERT and $0.30$ for SPLADE), but however to a lesser extent than BM25 ($\overline{\deltaRSJ}=0.83$). Finally, we observed that when transferring, all the models have a higher $\deltaRSJ$ variance compared to their trained version on MS MARCO: in all cases, the standard deviation (when normalized by BM25 one) is around $0.8$ for MS MARCO, but around $1.1$ for \texttt{TREC-COVID} and \texttt{FiQA-2018}. This further strengthens our point on the issue of generalizing lexical matching to out-of-domain collections.

\section{Conclusion}

In this work, we analyzed how different neural IR models predict the importance of lexical matching for query terms. We proposed to use the Robertson-Sparck Jones (RSJ) weight as an appropriate measure to compare term importance w.r.t. the user and system relevance. We introduce a black box approach that enables a systematic comparison of different models w.r.t. term matching. We have also investigated the behavior of lexical matching in the zero-shot setting. Overall, we have shown that lexical matching properties are heavily influenced by the presence of the term in the training collection.
The rarer the term, the harder it is to find documents containing that term for most neural models. Furthermore, this phenomenon is amplified if term statistics change across collections. 
\newpage




%
%


\bibliographystyle{splncs04}
\bibliography{ecir2022}

\end{document}